\documentclass[aps,prd,amsmath,twocolumn,longbibliography]{revtex4-1}

\usepackage{graphicx}\usepackage{multirow}  
\usepackage{amssymb,color}
\renewcommand{\bar}[1]{\overline{#1}}
\renewcommand{\bar}[1]{\overline{#1}}

\def\Dslash{\raise.15ex\hbox{/}\kern-.7em D}
\def\Pslash{\raise.15ex\hbox{/}\kern-.7em P}

\newcommand{\la}{\langle}
\newcommand{\ra}{\rangle}

\newcommand{\ben}{\begin{displaymath}}
\newcommand{\een}{\end{displaymath}}
\newcommand{\be}{\begin{equation}}
\newcommand{\ee}{\end{equation}}
\newcommand{\bwt}{\begin{widetext}}
\newcommand{\ewt}{\end{widetext}}
\newcommand{\bea}{\begin{eqnarray}}
\newcommand{\eea}{\end{eqnarray}}

\newcommand{\eq}[1]{Eq.~(\ref{#1})}

\def\g{\gamma}\def\a{\alpha}\def\b{\beta}

\newcommand{\beqn}{\begin{equation}}
\newcommand{\eeqn}{\end{equation}}

 \def\D{\Delta}

\newcommand{\<}{\left\langle}
\renewcommand{\>}{\right\rangle}

\begin{document}

\title{Charge Symmetry Breaking Effects on Neutron Beta Decay in  Non-Relativistic Quark Models }

\author{Jacob W. Crawford and Gerald A. Miller}

\affiliation{ 
Department of Physics,
University of Washington, Seattle, WA 98195-1560, USA}

\date{\today}

\begin{abstract}
A formalism for the study of charge symmetry breaking (CSB) effects is discussed and
 used to analyze the effects of charge symmetry breaking on  neutron beta decay.
The effect of including CSB  reduces the beta decay matrix element by an amount on the order of $10^{-4}$, a value  much larger than the previous estimate.
A much smaller contribution due to proton recoil is treated as well, which is found to be on the order of $10^{-6}$. 
The current uncertainty in the value of $V_{ud}$ is also of order $10^{-4}$.  An improvement of that uncertainty by an order of magnitude would require that charge symmetry breaking effects  should be included in future analyses.
 \end{abstract}  
\maketitle     
\noindent

\section{Introduction}

There is a great modern interest in  precisely determining  the CKM matrix elements.
This is particularly true of $V_{ud}$, since it provides the greatest contribution to the unitary condition $|V_{ud}|^2+|V_{us}|^2+|V_{ub}|^2=1$ which is a testing ground for searches for physics beyond the standard model (BSM).
The uncertainty of this condition comes in comparable parts from $V_{ud},V_{us}$~\cite{Cirigliano:2019wao}, meaning stronger statements about the possibility of a nonunitary CKM matrix can be made by reducing the uncertainty of $V_{ud}$.
Of particular interest to this work are the measurements of $V_{ud}$ via neutron beta decay.
A benefit of these experiments is a lack of nuclear structure dependent corrections that  add theoretical uncertainty to a measurement.
So the uncertainty on these experiments are largely experimental in nature.
This means of measuring $V_{ud}$ is expected to reach levels of precision in competition with superallowed decay experiments within the next decade~\cite{Cirigliano:2019wao}.

The need for greater precision impels us to re-examine the connection between the standard model Lagrangian expressed in terms of quarks and neutron beta decay. The implicit assumption is that the  quark-level isospin operator is the same as the nucleon isospin. operator. This is only true if the nucleon wave function is invariant under the isospin rotation known as the charge-symmetry rotation.
The accuracy of  the implicit assumption was examined by Behrends \& Sirlin\cite{old} who found, using an order-of-magnitude estimate, very small corrections on the order of 
10$^{-6}$.  The present paper is aimed at providing a more detailed estimate.


 Neglecting the mass difference and electromagnetic effects of the up and down quark leads to an invariance in the QCD Lagrangian under the interchange of up and down quarks. 
This invariance is called charge symmetry, which is a more restrictive symmetry than isospin.
The fundamental reason why the neutron is more massive than the proton is the fact that the down quark is more massive than the up quark. This positive contribution to $M_n-M_p$ is tempered by electromagnetic effects and the influence of quark masses on the one-gluon exchange potential, see the 
 reviews \cite{review1,review2,review3}. The influence of charge-symmetry breaking operators on the proton wave function and the resulting electromagnetic form factors were discussed in Ref.~\cite{main}.

 Here is an  outline. 
 Section II  introduces the necessary definitions and perturbation theory that is  the basis for our understanding of the  weak operator and charge symmetry. The non-relativistic quark model and the charge symmetry breaking interactions are discussed in Section III.
This Section includes the explicit definitions of our Hamiltonian, charge symmetry breaking operators, and all nucleon states  used in this work. The CSB effects are the mass difference between up and down quarks,  its influence in the kinetic energy and one-gluon exchange operators and electromagnetic effects. The parameters of the models are determined by the need to reproduce the measured mass difference between the neutron (n) and proton (p) in 
in Section IV. 
Section  V displays  the relevant perturbation theory. Evaluations are performed in Sect. VI, and the results are interpreted in
Section VII. 

\section{Formalism}

The weak operator which dictates the decay $n\to p+e+\overline{\nu}_e$ can be written
in terms of operators acting on quarks using first-quantized notation as
\begin{equation}
\label{eqn:Hw}
    H_w:=V_{ud}\sum_{i=1}^3\tau_+(i)\equiv V_{ud}\tau_+
\end{equation}
where $\tau_+|d\rangle=|u\rangle$ and $\tau_+|u\rangle=0$. The property is a statement that the $u,d$ system is a fundamental  isospin doublet. If the same is true of the neutron, proton system we may state that $\tau_+|n\rangle=|p\rangle$, so that $\<p|H_w|n\>=V_{ud}$. This expression is  used in all analysis that aim to extract the value of $V_{ud}$.

However, the neutron and proton are composite particles. The $u$ and $d$ quarks within have different masses and undergo electromagnetic interactions. Thus the expression $\tau_+|n\rangle=|p\rangle$, must be modified. 
It's necessary to introduce the isospin formalism\cite{review1}  to understand the modifications

The isospin rotation known as the charge  symmetry rotation operator is used to obtain the nucleon matrix element.
The logic is as follows.
The charge symmetry operator is the 180$^\circ$ isospin rotation operator defined~as
\begin{equation}
    P_{cs}=e^{i\pi T_2}
\end{equation}
with
\begin{equation}
    P_{cs}^\dagger uP_{cs}=d
\end{equation}
and
\begin{equation}
    T_2=\frac{1}{2}
    {q^\dagger}\tau_2q
\end{equation}
where $q=u,d$ is the quark-field operator.

If charge symmetry holds the neutron is obtained from the proton by the $P_{cs}$ isospin rotation. However,  
the Hamiltonian, $H$, can expressed in terms of a charge symmetry conserving term $H_0$ and a breaking term $H_1$ with $H=H_0+H_1$, and
$[H_0,P_{cs}]=0$. The eigenstates of $H_0$ are denoted in a round bracket notation: $|\cdots)$ states, and $|\cdots\rangle$ is used to denote the  physical eigenstates of  $H$.   
Then
 \begin{equation}
   H_0|p,m_s)=\sqrt{\overline{M}^2+\vec p^2}|p,m_s) ,
\end{equation}
with the label $p,m_s$ representing a proton of momentum $\vec p$ of spin $m_s$.
We  treat the physical wave function using first-order perturbation theory in  $H_1$:
\begin{equation}
    |p,m_s\rangle\approx\sqrt{Z}|p,m_s)+\frac{1}{\overline{M}-H_0}\Lambda H_1|p,m_s)
\label{fo}
\end{equation}
where the projection operator $\Lambda$ defined by
\begin{equation}
    \Lambda:=1-|p,m_s)(p,m_s|-|n,m_s)(n,m_s|
\end{equation}
 projects out the ground state degrees of freedom. The normalization factor
 $Z$ is  defined so that
\begin{equation}
    1=Z+(p,m_s|H_1\frac{\Lambda}{(\overline{M}-H_0)^2}H_1|p,m_s).
    \label{Zdef}
\end{equation}
The expression \eq{fo} is sufficient to account for terms of order $H_1^2$ in the beta decay matrix element.
Second order terms in the wave function lead to higher order contributions to the matrix element.

Using charge symmetry, the neutron and proton states obey the relation 
\begin{equation}
    |n,m_s)=P_{cs}|p,m_s).
    \label{ndef}
\end{equation}
The charge symmetry breaking piece of our Hamiltonian is $H_1$, and in first-quantized notation contains  operators $\tau_3(i)$, where $i$ labels a quark.  
The use of the identity
\begin{equation}
    P_{cs}^\dagger \tau_3(i)P_{cs}=-\tau_3(i).
\end{equation}
along with \eq{Zdef} and \eq{ndef} informs us that $Z$ of the neutron is the same as the $Z$ of the proton.

It is also useful to define the quantity 
\bea
   & \Delta H:=P_{cs}^\dagger HP_{cs}-H\label{DH}\\&=P_{cs}^\dagger H_1P_{cs}-H_1=-2H_1.
\eea
The  relation  
\bea
   (p|\Delta H|p)=(p|P_{cs}^\dagger HP_{cs}|p)-(p|H|p) \nonumber\\
   =(n|H|n)-(p|H|p)=M_n-M_p
    \label{eq:mass diff}
\eea
 will be used to fix the model parameters in Section~IV.


\section{Non-relativistic Quark Model}

In non-relativistic quark models the spin and momentum of the proton are unrelated, so we can write our proton state as
   $ |p,i)\to|p,\uparrow)
$
for a spin up proton. The spin index will be treated implicitly so that $|p,\uparrow)\to|p)$.

The Hamiltonian is specified by the terms
\begin{equation}
    H=K+V_{con}+V_{em}+V_{g}
\end{equation}
which are the kinetic energy $K$, confining potential $V_{con}$ (which respects charge symmetry), electromagnetic $V_{em}$ and gluon exchange $V_g$ interactions.
The charge symmetry breaking part of the Hamiltonian is then

\begin{equation}
\D H=    \Delta K+\Delta V_{em}+\Delta V_g,
\end{equation}
with each of the terms defined as in \eq{DH}.

We proceed to determine the individual contributions to $\D H$.  The first step is to  define the quark masses: $m_i=\overline{m}+\frac{\Delta m}{2}\tau_3(i)$ and $\Delta m=m_u-m_d$.
Then the non-relativistic kinetic energy \cite{hyperfine} 
is given 
\begin{equation}
    K=\sum_i(m_i+\frac{p_i^2}{2m_i}),
\end{equation}
and
\begin{equation}
    \Delta K=\Delta m\sum_i\tau_3(i)+\frac{\Delta m}{\overline{m}}\sum_i\frac{p_i^2}{2\overline{m}}\tau_3(i).
\end{equation}
The first term of $\Delta K$ does not contribute to any excitations, and may be  neglected in the calculation of beta decay amplitude. 

\begin{widetext}
The electromagnetic interaction is given by~\cite{coulomb}
\begin{equation}
    V_{em}=\alpha \sum_{i<j}q_iq_j(\frac{1}{r_{ij}}-\frac{\pi}{2\overline{m}^2}\delta(\vec r_{ij})[\frac{2}{\overline{m}^2}+\frac{4}{3}\frac{\vec{\sigma}(i)\cdot\vec{\sigma}(j)}{\overline{m}^2}])
\end{equation}
where $q_i=\frac{1}{6}+\frac{1}{2}\tau_3(i)$ and $r_{ij}=|\vec{r}_i-\vec{r}_j|$.
The charge asymmetric contribution from this operator is~\cite{main}
\begin{equation}
    \Delta V_{em}=-\frac{\alpha}{6}\sum_{i<j}[\tau_3(i)+\tau_3(j)](\frac{1}{r_{ij}}-\frac{\pi}{2\overline{m}^2}\delta(\vec r_{ij})[1+\frac{2}{3}\vec{\sigma}(i)\cdot\vec{\sigma}(j)]).
\end{equation}
\end{widetext}
The gluon exchange operator is taken to be
\begin{equation}
    V_g=-\alpha_s\sum_{i<j}\lambda_i\cdot\lambda_j[\frac{\pi}{2}\delta(\vec r_{ij})(\frac{1}{m_i^2}+\frac{1}{m_j^2}+\frac{4}{3}\frac{\vec{\sigma}(i)\cdot\vec{\sigma}(j)}{m_im_j})]
\end{equation}
where for three quark baryons, $\lambda_i\cdot\lambda_j=-\frac{2}{3}$\cite{hyperfine,coulomb}.
The long range $1/r_{ij}$ term of $V_g$ respects charge symmetry, and so is not included.
The charge symmetry breaking piece of this interaction is given by\cite{main}

\begin{equation}
    \Delta V_g=\alpha_s\frac{2\pi}{3}\frac{\Delta m}{\overline{m}^3}\sum_{i<j}[\tau_3(i)+\tau_3(j)]\delta(\vec r_{ij})[1+\frac{2}{3}\vec{\sigma}(i)\cdot\vec{\sigma}(j)].
\end{equation}
We use an SU(6) space-spin-wave-function along with  oscillator confinement to represent the charge-symmetric wave functions.
Then  we  write
\begin{equation}
    |p)=|\psi_0\rangle\frac{1}{\sqrt{2}}(|\phi_s\rangle|\chi_s\rangle+|\phi_a\rangle|\chi_a\rangle)
\end{equation}
where
\begin{equation}
    \<\vec r_i|\psi_0\>=\psi_0(\rho,\lambda)=Ne^{-\frac{\rho^2+\lambda^2}{2\beta^2}}.
\end{equation}
Here $\vec\rho=\frac{1}{\sqrt{2}}(\vec r_1-\vec r_2)$ and $\vec\lambda=\frac{1}{\sqrt{6}}(\vec r_1+\vec r_2-2\vec r_3)$, and center of mass dependence is not made explicit.
The standard mixed symmetry flavor ($\phi_{s,a})$ and spin $(\chi_{s,a}$) wave functions are used\cite{quarkspartons}.
\section{Model Parameters}

The parameters of the non-relativistic quark model shall be determined from the neutron proton mass difference and a consideration of pionic effects. These parameters are $\b,\a_s$ and $\bar m$, and these are constrained by values of the proton's charge radius, magnetic moment and the $\D-$nucleon mass splitting. The model does not include the explicit effects of the pion cloud because those are charge symmetric if the pion-nucleon coupling constant is taken (consistent with observations) to be charge symmetric. However, any consideration of the values of parameters must take implicit account of the pion cloud.  Here we follow the ideas of the Cloudy Bag Model \cite{Theberge:1980ye,Thomas:1981vc,Theberge:1981mq} in which a perturbative treatment of pions as quantum fluctuations converges for bag radii (confinement radius) greater than about 0.6 fm. The importance of pionic effects decreases as the confinement radius of the model increases. The effects of the pion cloud contribute to the magnetic moment and to the $\D-$nucleon mass splitting.

The quark contribution to the root-mean-square charge radius is  $\b$. The measured value is $0.84$ fm. The proton magnetic moment 
is 2.79 nm, and the gluon-exchange contribution ($g$) to the  $\D-$nucleon mass splitting of about 303 MeV is given by
\bea
(M_\D-M_N)_g= {2\over 3}\sqrt{2\over \pi }{\a_s\over {\bar m}^2\b^3}.
\eea

We use three separate models to evaluate the effects of charge symmetry breaking on beta decay. We start with a large confinement radius of $\b=0.837$ fm, with small pionic effects so that $\g$ the fraction of the  $\D-$nucleon mass splitting is large, 0.9 and the
quark mass is taken to be 2 fm$^{-1}$ accounting for about 084\% of the proton magnetic moment with the pion-cloud accounting for the remainder. The other two models are obtained by increasing the value of $\bar m$, decreasing the value of  $\b$, thus  decreasing the value of $\gamma$.
Then the value of $\D m$ is chosen so that  according to \eq{eq:mass diff}:
\begin{equation}
    (p|\Delta H|p)=M_n-M_p\approx1.29\text{ MeV}.
\end{equation}


Evaluating the individual terms yields

\begin{equation}
    (p|\Delta K|p)=\Delta m\left(\frac{1}{2\overline{m}^2\beta^2}-1\right)
\end{equation}

\begin{equation}
    (p|\Delta V_{em}|p)=-\frac{\alpha}{3\beta}\sqrt{\frac{2}{\pi}}\left(1-\frac{5}{12\overline{m}^2\beta^2}\right)
\end{equation}

\begin{equation}
    (p|\Delta V_g|p)=\frac{5\alpha_s\Delta m}{9\overline{m}^3\beta^3}\sqrt{\frac{2}{\pi}}.
\end{equation}
Then the sum of each term yields approximately the desired mass difference in each non-relativistic quark model.

The parameter values of each  our models can be found in Table \ref{tab:parameters}.
\begin{table}[h]
    \centering
    \begin{tabular}{llllll}
    \toprule
    \multicolumn{6}{c}{Model Parameters}\\
  Model  & $\beta^2$ (fm$^{2}$) & $\overline{m}$ (fm$^{-1}$) & $\alpha_s$ & $\Delta m$ (MeV) & $\gamma$\\
    1 & 0.7 & 2 & 6.1 & -6.9 & 0.9\\
    2 & 0.6 & 2.1 & 4.7 & -5.3 & 0.8\\
    3 & 0.5 & 2.2 & 3.5 & -4.5 & 0.7\\
  \hline
\end{tabular}
    \caption{Model parameters adjusted from Ref. \cite{main}}
    \label{tab:parameters}
\end{table}

\section{beta decay matrix element}
We compute the  matrix element of $\tau_+$ using the perturbed state of \eq{fo} and the related one for the neutron. The result is
\bea&
    \langle p|\tau_+|n\rangle= 
     Z+( p|H_1\frac{\Lambda}{\overline{M}-H_0}\tau_+\frac{\Lambda}{\overline{M}-H_0}H_1|n),
\label{me}\eea
in which the terms of  first order  in $H_1$ above vanish because the resolvent $\frac{\Lambda}{\overline{M}-H_0}$ has the states $|p),|n)$ projected out, and $\tau_+$ can only take nucleons to other nucleons. Moreover the matrix element of $\tau_+$ between the bare 
neutron and state is unity. 

The operator  $H_1$ conserves spin angular momentum, so that $(\Delta|H_1|N)=0$, and there is no contribution due to $\Delta$ baryons.
There is also no way for $H_1$ to mix in states containing strange or heavy quarks, so the only contributions are due to spatial excitations of nucleons.
Further, recall that $\tau_+|n)=|p)$ and $\tau_+|p)=0$, and that $H_1$ introduces no units of angular momentum, so the only excitations can be s-waves.
This means that the intermediate states appearing in \eq{me} and \eq{Zdef} are of the form  form $|n^*)( n^*|$ of  $|p^*)(p^*|$, in which 
the $^*$ notation refers to radial excitations.

With this notation for radial excitations the normalization factor $Z$ of \eq{Zdef} can be written as
\bea
    Z=1-\sum_{k\neq0}\frac{\langle p|H_1|p^*_k\rangle\langle p^*_k|H_1|p\rangle}{(\overline{M}-M_k)^2}\\
    =1+\sum_{k\neq0}\frac{\langle p|H_1|p^*_k\rangle\langle n^*_k|H_1|n\rangle}{(\overline{M}-M_k)^2}
\eea

The second term of this equation is equal to the second term of \eq{me}.
 The net result is that 
 \begin{equation}
\langle p|\tau_+|n\ra=Z +    \sum_{k\neq0}\frac{( p|H_1|p^*_k)( n^*_k|H_1|n)}{(\overline{M}-M_k)^2}.
\label{r}\end{equation}
The deviation of $Z$ from unity is the same as that seen in the second term of the above equation.
Note that the correction to unity is negative because the neutron and proton matrix elements have opposite signs.
\section{Evaluation}

We  next compute the individual contributions to the correction.
This will first be done using  the assumption that the proton is stationary after the decay, after which we include a nonzero momentum transfer.

We  use  the notation
\begin{equation}
    |p_k^*)=|\psi_k\rangle\frac{1}{\sqrt{2}}(|\phi_s\rangle|\chi_s\rangle+|\phi_a\rangle|\chi_a\rangle)
\end{equation}
to reference the $k$th radial excitation, where
\begin{equation}
    \<{\rho}|\psi_k\>=R_{k0}(\rho):=\sqrt{\frac{2(k!)}{\beta^3\Gamma(k+\frac{3}{2})}}\exp\left(\frac{-\rho^2}{2\beta^2}\right)L_k^{1/2}\left(\frac{\rho^2}{\beta^2}\right)
\end{equation}
is the radial wave function   
Here, $L^{1/2}_k$ is a generalized Laguerre Polynomial\cite{structure}.
So the quantity to be calculated is  the second term of \eq{r}.
 and the mass denominator can be written
\begin{equation}
    M_k-\overline{M}=\frac{2k}{\overline{m}\beta^2},
\end{equation}
which is just the energy added due to the harmonic oscillator excitation.


\subsection{Zero Recoil}

We will first turn our attention to the electromagnetic interaction matrix element:
\bwt
\bea&
    ( p_k^*|\Delta V_{em}|p)
=-\frac{\alpha}{2}( p_k^*|[\tau_3(1)+\tau_3(2)](\frac{1}{\sqrt{2}}\frac{1}{\rho}-\frac{\pi}{\overline{m}^2\sqrt{2}}\delta(\vec \rho)[1+\frac{2}{3}\vec{\sigma}(1)\cdot\vec{\sigma}(2)])|p)\nonumber\\&
    =-\frac{\alpha}{2\sqrt{2}}\frac{2}{3}((\psi_k|\frac{1}{\rho}|\psi_0)
-\frac{\pi}{\overline{m}^2}\frac{5}{3}(\psi_k|\delta(\vec\rho)|\psi_0)).
\eea
Then the gluon exchange term is calculated analogously to the electromagnetic contact term, so we can simply write
\begin{equation}
    (p_k^*|\Delta V_g|p)
=\frac{20\pi\alpha_s}{9\sqrt{2}}\frac{\Delta m}{\overline{m}^3}(\psi_k|\delta(\rho)|\psi_0).
\end{equation}
Lastly, we must calculate the kinetic energy term,
\begin{equation}
    (p_k^*|\Delta K|p)
=\frac{\Delta m}{3\overline{m}^2}(\psi_k|p_\rho^2|\psi_0)
\end{equation}
The necessary integrals to complete the above expressions can be found in Table \ref{tab:Integrals}.

\begin{table}[h]
    \centering
    \begin{tabular}{llll}
 \hline\hline
    \multicolumn{4}{c}{Relevant Integrals}\\
   \hline
    $(\psi_k|\frac{1}{\rho}|\psi_0)$ &&&  $\frac{\Gamma(k+\frac{1}{2})}{\beta\sqrt{\pi}}\sqrt{\frac{2}{k!\sqrt{\pi}\Gamma(k+\frac{3}{2})}}$\\
    \\
    $(\psi_k|\delta(\vec\rho)|\psi_0)$ &&& $\frac{1}{\pi^{3/2}\beta^3}\sqrt{\frac{2\Gamma(k+\frac{3}{2})}{k!\sqrt{\pi}}}$\\
    \\
    $(\psi_k|p_\rho^2|\psi_0)$ &&& $\frac{3}{4\beta^2}\sqrt{\frac{\pi}{\Gamma(\frac{3}{2})\Gamma(k+\frac{3}{2})}}(\delta_{k0}+\delta_{k1})$\\
    \\
    $\frac{\sqrt{6}}{q\beta^3}\int\lambda d\lambda \sin\left(\sqrt{\frac{2}{3}}q\lambda\right)\left(\frac{\lambda^2}{\beta^2}\right)^{i+l}e^{-\frac{\lambda^2}{\beta^2}}$ &&& $\Gamma(i+l+\frac{3}{2})F_1^{(1)}(i+l+\frac{3}{2},\frac{3}{2},-\frac{Q^2\beta^2}{6})$\\
 \hline
    \end{tabular}
    \caption{The function $F_1^{(1)}$ is the confluent hypergeometric function of the first kind.}
    \label{tab:Integrals}
\end{table}

 \ewt
 
 The series of \eq{r} is evaluated simply by taking the sum of all of the terms, so we explain why the series converges.
 The large-$k$ values of the contact potential are controlled by the factor
\begin{equation}
    \sqrt{\frac{\Gamma(k+3/2)}{k!}}\sim k^{1/4}.
\end{equation}
This function increases without bound, but the mass difference in the denominator is linear with $k$, so the series is convergent.
The factor $\frac{\Gamma(k+\frac{1}{2})}{\beta\sqrt{\pi}}\sqrt{\frac{2}{k!\sqrt{\pi}\Gamma(k+\frac{3}{2})}}$  associated with the Coulomb falls as $1/k^{3/4}$ so it  yields a convergent series. The kinetic energy term only enters for $k=1$. The net result is that  the perturbation series is convergent.

\begin{table}[h]
    \centering
    \begin{tabular}{llll}
    \toprule
    \multicolumn{4}{c}{Model 1}\\
    Source & $(0|\Delta H|0)$ (MeV) & $(1|\Delta H|0)$ (MeV) & Importance \\
    $\Delta K$ & 5.66786 & -1.00604 & Second\\
    $\Delta V_{em}$ & -0.38973 & -0.10347 & Least\\
    $\Delta V_g$ & -3.98475 & -4.88030 & Most\\
    \toprule
    \multicolumn{4}{c}{Model 2}\\
    $\Delta K$ & 4.29849 & -0.81773 & Second\\
    $\Delta V_{em}$ & -0.41667 & -0.10652 & Least\\
    $\Delta V_g$ & -2.59111 & -3.17345 & Most\\
    \toprule
    \multicolumn{4}{c}{Model 3}\\
    $\Delta K$ & 3.57025 & -0.75914 & Second\\
    $\Delta V_{em}$ & -0.44847 & -0.10693 & Least\\
    $\Delta V_g$ & -1.83750 & -2.25047 & Most\\
\end{tabular}
    \caption{Results for the three models of Table I  are presented. The first column shows the ground state matrix element of the different contributions to $\Delta H=-2H_1$. The second column shows the connection between the ground state $|0)$ and the first excited state $|1)$.  The column labelled ``importance" assesses the importance of each term as determined by setting each individually to zero and checking the change in the total correction.} 
    \label{tab:contr}
\end{table}

The significance of each contribution is listed obtained with different  models is displayed in Table \ref{tab:contr}. The  kinetic energy term which contains the effect of the mass difference between up and down quarks controls the sign of the neutron-proton mass difference. The relative importance of each term of $\D H$ is also displayed and depends on the value of $\b$. It is noteworthy that there are cancellations between the separate terms of $\D H$ that result in the n-p mass difference of 1.29 MeV for each model. However, the contributions  of the separate terms  to the dominant matrix element  $(1|\Delta H|0)$ all have the same sign. This is mainly because the quark mass difference term does not  involve the spatial wave function an so cannot convert the ground state to any excited state. Note especially that the sum of the three terms is much larger than the individual terms and  the square that enters in computing the beta decay matrix element.

In particular the ratio, $R$ given by
\bea 
R\equiv \left((1|\Delta H|0)\over (0|\Delta H|0)\right)^2
\label{rat}\eea
 varies between about 6 and 22 as one changes the models from 3 to 1.

\subsection{Nonzero Proton Recoil}

The next step is to endow the Hamiltonian $H_w$ to take the momentum transfer $\vec q$ to the final proton into account. 
This is done by making the operator substitution $\tau_+\to\tau_+^*(\vec q)$:
\begin{equation}
    \tau_+^*(\vec q)=\sum_{j}\tau_+(j)e^{i\vec q\cdot\vec r_j}.
\end{equation}
The second order quantity to be calculated now is the matrix element $\la \vec q|\tau_+^*|{\vec 0}\ra $ with
\bea 
&\la \vec q |\tau_+^*|{\vec 0}\ra \equiv 
    Z(p|e^{-i\sqrt{\frac{2}{3}}\vec{q}\cdot\vec{\lambda}}|n)\nonumber\\&+
    \frac{1}{4}\sum_{j,k\neq0}\frac{(p|\Delta H|p_j^*)(p_j^*|e^{-i\sqrt{\frac{2}{3}}\vec{q}\cdot\vec{\lambda}}|p_k^*)(n_k^*|\Delta H|n)}{(\overline{M}-M_j)(\overline{M}-M_k)}.
\label{rc}\eea 
The correction to the beta decay matrix element is defined to be $\D(|\vec q|)>0$ with
\bea 
\la \vec q |\tau_+^*|{\vec 0}\ra =1-\D(|\vec q|)
.\eea

Before calculating the remaining matrix element, we need the average momentum transfer to the proton during beta decay.
This is accomplished by using the recoil spectrum found on Pg. 14 of the PhD thesis of G. Konrad\cite{protonspec}, originally derived by O. Nachtmann\cite{origprot}.
The spectrum and related functions are written:
\begin{equation}
    w_p(T)\propto g_1(T)+ag_2(T)
\end{equation}
\begin{widetext}
\begin{equation}
    g_1(T)=\left(1-\frac{x^2}{\sigma(T)}\right)^2\sqrt{1-\sigma(T)}\left[4\left(1+\frac{x^2}{\sigma(T)}\right)-\frac{4}{3}\left(1-\frac{x^2}{\sigma(T)}\right)\left(1-\sigma(T)\right)\right]
\end{equation}

\begin{equation}
    g_2(T)=\left(1-\frac{x^2}{\sigma(T)}\right)^2\sqrt{1-\sigma(T)}\left[4\left(1+\frac{x^2}{\sigma(T)}-2\sigma(T)\right)-\frac{4}{3}\left(1-\frac{x^2}{\sigma(T)}\right)\left(1-\sigma(T)\right)\right]
\end{equation}
\end{widetext}
\begin{equation}
    \sigma(T)=1-\frac{2TM_n}{\Delta^2}
\end{equation}

\begin{equation}
    x=\frac{m_e}{\Delta},\quad\Delta=M_n-M_p=1293.333(33)~\rm keV.
\end{equation}
Note that this spectrum does not take into account Coulomb or radiative corrections, but that level of precision is not necessary for the current application.
We use the first moment of a normalized $w_p$ to get the mean kinetic energy of the recoiled proton:
\begin{equation}
    \<T\>=\frac{\int Tw_p(T)dT}{\int w_pdT}=357.177\text{ eV}
\end{equation}
where the domain of the given integral was taken as $(0,751$ eV).
This can then be converted into the average wave number, $ \<Q\>$, of the recoiled proton
\begin{equation}
    \<Q\>=\frac{\sqrt{2M_p\<T\>}}{\hbar c}=4.1\times10^{-3}\text{ fm}^{-1}.
\label{Q}\end{equation}
We use $|\vec q|=\la Q\ra$ in the following calculations

The  effect of a non-zero value of $\<Q\>$ in the first term of \eq{Q} is given by the deviation between the factor 
$e^{-\frac{Q^2 \beta ^2}{6 }}$ and unity, which is of order of magnitude 10$^{-6}$.
In order to compute the analytic expression for the matrix element $\<\psi_j|e^{-i\sqrt{\frac{2}{3}}\vec q\cdot\vec\lambda}|\psi_k\>$, the closed form for the Laguerre Polynomials,
\begin{equation}
    L_j^{(\alpha)}(x)=\sum_{i=0}^j\frac{(-1)^i}{i!}\binom{j+\alpha}{j-i}x^i
    \label{eq:laguerre}
\end{equation}
must be used\cite{AnS}.
The final result contains a double sum due to (\ref{eq:laguerre}) over the final integrated expression found in Table \ref{tab:Integrals}.
The second term of \eq{rc}   contributes only at the level of $10^{-10}$.

The numerical results of this and the preceding section can be found in Table \ref{tab:results}.

\begin{table}[h]
    \centering
    \begin{tabular}{ccc}
    \toprule
    \multicolumn{3}{c}{Results}\\
       Model  &$\D(0)$ $(\times10^{-4})$ &$\D(\<Q\>)$ $(\times10^{-4})$ \\
        1 & 4.0297 & 4.0101\\
        2 & 1.4668 & 1.4500 \\
        3 & 0.6146 & 0.6005\\
    \end{tabular}
    \caption{Computed changes to the value of the beta decay matrix element caused by charge symmetry breaking effects.} 
    \label{tab:results}
\end{table}

For all of the models the change in the beta decay matrix element is  a reduction of the order of $10^{-4}$. 
 Table III shows us that this change is about an order of magnitude larger than that caused by any one of the terms of $\D H$. 
The model dependence arises from the different values of the length parameter and quark masses.
These affect the energy denominator.
The table shows that the effect of including the non-zero value of the momentum transfer is of the presently negligible order of $10^{-6}$, 

The current value~\cite{Workman:2022ynf} of $V_{ud}$ is given by
 \bea |V_{ud}| = 0.97373 \pm  0.00031, \eea
so the size of the charge symmetry breaking effect is of the order of the current uncertainty.
\section{Discussion and Assessment}

A general formalism for including  effects of charge symmetry breaking (CSB) is discussed and applied to computing  neutron
 beta decay matrix elements. 
CSB effects are  known to enter only at second and higher order~\cite{old}. 
The known CSB effects are the quark mass differences, the effect of quark mass differences on the kinetic energy and gluon exchange potentials and electromagnetic effects.
These are evaluated using
 three non-relativistic quark models  using of oscillator confinement.
Our second-order result is that including CSB effects
reduce the beta decay matrix element by about  $10^{-4}$. Thus higher orders need not be included.
The calculations involved summing over many intermediate states, but the dominant terms  arise from including the first radial excitation.
 This task was made simpler by the acquisition of analytic results for each matrix element, taking as many terms as necessary for a sufficiently converged result.
Three non-relativistic quark models were compared in the analysis,

Calculations were done with and without including the effects of proton recoil \cite{protonspec}. The latter effect is of order $10^{-6}$ and currently negligible, justifying that the proton can be considered a body at rest in the context of neutron beta decay.

It is interesting that our result is about 100 times larger than that of the original 
 work of Behrends \& Sirlin~\cite{old}. There it was predicted that effects due to charge symmetry breaking on neutron beta decay should be on the order $10^{-6}$.
Their schematic calculation correctly used the square of the ratio of a matrix element divided by an energy denominator:  $\left(\frac{(p|H_1|p)}{\overline{M}}\right)^2\approx\left(\frac{1.3}{940}\right)^2\approx2\times10^{-6}$. They used the n-p mass difference as a matrix element instead of   the matrix element of  sum of the individual terms between the ground and excited states. While the individual terms tend to cancel in computing the n-p difference, they add coherently in computing the excitation matrix elements. This gives rise to enhancements of between about 6 and 21, as seen in \eq{rat}. Furthermore, the relevant  energy denominator is not the nucleon mass, but the excitation energy which is about half of that. Our  lowest and most important  energy denominator $\Delta M=\frac{2(\hbar c)^2}{\overline{m}\beta^2}$ varies between 280 and 360  MeV. The values, determined by using the correct approximate size of the nucleon,  are lower than the 500 MeV difference between the nucleon mass and its first excited state.  This reflects a long-standing problem of the non-relativistic quark model. 

The value of $\b^2$ could be decreased by a factor of about 50-60\% to increase the energy difference to about 500 MeV, but the matrix element of, for example, the gluon exchange term (which is the most important one for each of the models) varies as  $1/\b^3$ so the net result would be an increase the size of the CSB effect by $1/\b^2$, an increase of 50-60 \%. Thus we regard the results in
Tables III and IV. to be reasonable first semi-realistic estimates.

 We summarize by saying that the size of the CSB effects are to decrease the value of the beta decay matrix element by a factor of about $10^{-4}$, which is corroborated by an earlier Bag Model calculation found in Ref. \cite{Guichon_2011}. This is of the order of the current uncertainty. An improvement of that uncertainty by an order of magnitude would require that  charge symmetry breaking effects be included in future analyses. 

\section*{Acknowledgements}
This work was supported by the U. S. Department of Energy Office of Science, Office of Nuclear Physics under Award Number DE-FG02-97ER-41014.

\bibliography{biblio}

  \end{document}